\documentclass[conference]{IEEEtran} 
\usepackage{enumerate}
\usepackage{tablefootnote}
\usepackage{url}
\usepackage{balance} 

\begin{document}

\hyphenation{Oka-nagan pro-ject pro-jects cour-ses star-ted wi-de-ly achi-eved}

\title{Students Programming Competitions as an Educational Tool and a Motivational Incentive to Students}

\author{\IEEEauthorblockN{Youry Khmelevsky}
\IEEEauthorblockA{Computer Science, Okanagan College\\  Kelowna, BC Canada\\
Email: YKhmelevsky@okanagan.bc.ca}\\
\and
\IEEEauthorblockN{Ken Chidlow}
\IEEEauthorblockA{Computer Science, Okanagan College\\ Kelowna, BC Canada \\
Email: KChidlow@okanagan.bc.ca}
}

\maketitle

\begin{abstract}
In this short paper we report on student programming competition results by students from the Computer Science Department (COSC) of Okanagan College (OC) and discuss the achieved results from an educational point of view. We found that some freshmen and sophomore students in diploma and degree programs are very capable and eager to be involved in applied research projects as early as the second semester, and into local and international programming competitions as well. Our observation is based on the last 2 educational years, beginning 2015 when we introduced programming competitions to COSC students. Students reported that participation in competitions give them motivation to effectively learn in their programming courses, inspire them to learn deeper and more thoroughly, and help them achieve better results in their classes. 
\end{abstract}
%
%
%

%
%



\section{Introduction}

Since 2005 we have searched for new solutions and pedagogical approaches in COSC courses in our Computer Information Systems (CIS) diploma and Bachelor of Computer Information System (BCIS) degree programs. We had many department meetings to discuss students' performance and achievements, especially in COSC 111---Computer Programming I, COSC 121---Computer Programming II, COSC 224---Projects in Computer Science, COSC 470---Software Engineering and COSC 471---Software Engineering Project Project courses  \cite{Khmelevsky:2009:SDP:1536274.1536292, Khmelevsky:2012:ACG:2247569.2247578, Khmelevsky:2011:DLC:1989622.1989627, khmelevsky2015hybrid, Khmelevsky:2011:ICS:1989622.1989637, Khmelevsky:2011:RTS:1989622.1989638, khmelevsky2013strategies, Khmelevsky:2016:TYC:2910925.2910949, Khmelevsky:2016:NPT:2910925.2910937, 7237130}. We were concerned about an apparent low motivation factor in the students and high attrition rate within our institution's first year programming courses COSC 111/121. In an attempt to improve student motivation, starting from 2005 we introduced industry related capstone projects (projects with real industrial clients) within our COSC 224/236/470/471 courses and since 2007 we introduced applied research projects. Several local and international industrial clients supported our initiatives \cite{Khmelevsky:2016:TYC:2910925.2910949}, but it took almost 7 additional years to get support for our student research or research related projects from NSERC. Several industrial projects achieved success that we were able to obtain an NSERC Community College Innovation (CCI), Applied Research (ARD) and Engage College grants, two {\sc Mitacs} grants with SFU and UBCO, an Amazon's AWS in Education Research grant, and several small grants funded by the College for the student research projects with industrial sponsors. Despite our success in the second and fourth year courses, we still had attrition problems and some problems with low student motivation in our first year classes. 

A number of educational research papers on student competitions' design and evaluation are already published \cite{Lachi:2011:SC:2254516.2254537, 7084518, Harrell:2015:SCC:2831425.2831428}, but in our short paper here, we discuss the educational impact on introductory programming courses via organization of student competitions and student participation in annual programming competitions. 

In ``Student competitions as highly valuable learning experiences'' \cite{Topi:2013:SCH:2465085.2465093} H. Topi from Bentley University mentioned, that ``as faculty members who teach and design information systems (IS) courses and curricula, we frequently struggle to increase students' level of intellectual engagement and their motivation to focus on the material that we are covering.'' The author suggests ``to consider various forms of student competitions as a way to achieve a number of benefits for their students at a relatively low cost" for information systems (IS) students. He continues in \cite{Topi:2013:SCH:2465085.2465093} ``just the idea of participating in a competition is often enough to increase significantly students motivation level to learn and perform well." 

In ``A new paradigm for programming competitions" \cite{Bowring:2008:NPP:1352322.1352166} J.F Bowring tells that  ``faculty discussions have for some time focused on declining enrollments and on how to recruit students into computing''. They view their annual high school programming competition as a key component of their recruiting efforts.

In the next section we will discuss our competition results from an educational point of view. 

\section{Programming Competitions Introduction Results}

A couple of years ago several COSC freshmen students contacted us with a request to organize students programming competitions and to be accepted into our undergraduate research projects. They successfully competed in the ACM International Collegiate Programming Contest ICPC\footnote{\url{https://icpc.baylor.edu/}} and Institute of Electrical and Electronics Engineers (IEEE) IEEExtreme Programming world-wide competition\footnote{\url{http://tinyurl.com/28bol4x}} and achieved good results. 

In November 2015 our freshmen students participated in the fall's  IEEEXtreme. One of Okanagan College's first-year Bachelor of Computer Information Systems teams placed in the top 25 teams across Canada, and top 500 world-wide out of 2,000 global teams (and more than 6,400 students). This was the first year the College entered\footnote{\url{http://tinyurl.com/k4utmmg}}. In the following competition in 2016 our students achieved much better results: 2nd place in Canada and 125th of 2,200 teams globally at the IEEEXtreme 10th annual student programming competition\footnote{\url{http://tinyurl.com/ltk3n87}}. Students participated in the IBM/ACM ICPC contests as well, but in regional levels only. 

In 2017 a pair of OC computer science students logged an impressive showing at MIT's longest running programming competition ``Battlecode". This year's competition saw more than 1000 teams registered. In Okanagan College's very first appearance at the competition, two freshman students placed 49th overall, competing against top post-secondary institutions from all over the world\footnote{\url{http://tinyurl.com/kjxp3mm}}.

In the same timeframe, as an experiment, we accepted into our research projects a couple of freshmen students who participated in the programming competitions. We found that some talented freshman students can contribute to the research projects at the same level as, if not beyond, 3rd and 4th year undergraduate students. 

We continue organizing and supporting programming competitions for students who desire to compete with other universities and colleges in Canada as well as internationally. We discovered that our retention improved in the last 2 years and many more applicants enrolled and paid their tuition earlier in comparison with previous years. As of March 17th, 2017 73\% of reserved seats for the Computer Science department are already filled for the 2017-2018 educational year and more applications have been received compared to previous years.  

\section{CONCLUSION}
\label{Conclusion}

``Participation in student competitions can have valuable, highly important benefits for students, faculty members, and their educational institutions. Fortunately, there currently are several good national competitions for students in information systems programs" \cite{Topi:2013:SCH:2465085.2465093}. 

As a faculty member at Okanagan College mentioned, we have a new generation of students who are looking for challenges during their study at post-secondary institutions. They like to work hard, but they want to be rewarded for their hard work. Many students apply to the College because they prefer small class sizes and the lower tuition costs compared to universities. On the other hand they are demanding a high quality education, involvement in the worldwide programming competitions (IEEEXtreme, ICPC, and others), and involvement in industrial applied research projects, especially supported by Government grants and/or by large corporations. Many freshmen students want to continue their education at top universities within MSc and even in PhD programs. 

\section{ACKNOWLEDGMENTS}

We would like to thank NSERC for supporting five our applied research projects in 2014 - 2017. 

Our thanks to ``AWS Programs for Research and Education" program\footnote{\url{https://aws.amazon.com/grants/}} for supporting our research and educational projects, as well as to Atlassian. We thank ``IBM Academic Initiative" program\footnote{\url{https://developer.ibm.com/academic/}} for supporting us with Rational Rose engineering software, licenses and with training materials. 

Thanks to Jim Nastos for his valuable comments and edits towards the improvement of the readability of this paper.


%

%
\bibliographystyle{IEEEtran}
\balance
\bibliography{YouryWCCCE2017}  
%
%

\end{document}